\RequirePackage[l2tabu, orthodox]{nag}

\documentclass[10pt,journal,compsoc,final]{IEEEtran}

\usepackage[T1]{fontenc}                    
\usepackage[utf8]{inputenc}                 
\usepackage[english]{babel}

\usepackage[pdftex]{graphicx}               
\usepackage{listings}                       
\usepackage[nolist]{acronym}                
\usepackage[colorlinks]{hyperref}           
\usepackage[all]{hypcap}                    
\usepackage{cleveref}                       
\usepackage{microtype}                      

\usepackage[nocompress]{cite}

\usepackage[usenames,dvipsnames]{xcolor}
\definecolor{lst1}{RGB}{228,26,28}
\definecolor{lst2}{RGB}{55,126,184}
\definecolor{lst3}{RGB}{77,175,74}

\usepackage{float}
\newfloat{span}{tbp}{los}

\usepackage{pifont}
\newcommand{\cmark}{\ding{51}}

\makeatletter
\def\lst@makecaption{%
  \def\@captype{table}%
  \@makecaption
}
\makeatother

\usepackage[obeyFinal,textsize=footnotesize]{todonotes}
\usepackage{marginnote}

\usepackage{ifdraft}
\ifoptionfinal{}{
  \setlength{\paperwidth}{242mm}
  \setlength{\marginparwidth}{33mm}
  \setlength{\marginparsep}{7pt}
}

\makeatletter
\let\org@@cref\@cref
\renewcommand*{\@cref}[2]{%
  \edef\process@me{%
    \noexpand\org@@cref{#1}{\zap@space#2 \@empty}%
  }\process@me
}
\makeatother

\newcommand\code{\lstinline}

\ifoptionfinal{}{\overfullrule=2cm}

\usepackage{tikz}
\newcommand*\circled[1]{\tikz[baseline=(char.base)]{
            \node[shape=circle,draw,inner sep=.75pt] (char) {#1};}}

\usepackage{textcomp} 
\usepackage{siunitx}
\begin{document}

\lstset{%
  basicstyle          = \ttfamily%
                        \lst@ifdisplaystyle\scriptsize\fi,
  keywordstyle        = \color{lst2},
  stringstyle         = \color{lst1},
  commentstyle        = \itshape\color{lst3},
  showstringspaces    = false,
  frame               = top,
  frame               = bottom,
  framextopmargin     = 2pt,
  framexbottommargin  = 2pt,
  framexleftmargin    = 17pt,
  xleftmargin         = 17pt,
  belowskip           = 0ex,
  numbers             = left,
  numbersep           = 7pt,
  escapechar          = \¶
}

\lstdefinelanguage{CUDAC}[ANSI]{C}{%
  morekeywords={%
    __global__,__device__,%
    threadIdx,blockIdx,blockDim,gridDim},%
}


\lstdefinelanguage{Julia}{%
  morekeywords={%
    type,primitive,abstract,struct,new,%
    function,@generated,macro,module,where,%
    begin,end,do,%
    try,catch,return,%
    const,export,import,using,%
    if,elseif,else,for,while,break,continue,%
    true,false,quote},%
  sensitive=true,%
  alsoother={\$},%
  morecomment=[l]\#,%
  morecomment=[n]{\#=}{=\#},%
  morestring=[s]{"}{"},%
  morestring=[m]{'}{'},%
}[keywords,comments,strings]

\lstdefinelanguage{JuliaCUDA}[]{Julia}{%
  morekeywords={%
    @cuda,%
    CuArray,CuDeviceArray,%
    CuDevice,CuContext,destroy,%
    blockIdx,blockDim,threadIdx,threadDim,gridDim,warpsize},%
}

%
%

\begin{acronym}

\acro{api}[API]{Application Programming Interface}
\acro{ast}[AST]{Abstract Syntax Tree}
\acro{cuda}[CUDA]{Compute Unified Device Architecture}
\acro{dsl}[DSL]{Domain Specific Language}
\acro{ffi}[FFI]{Foreign Function Interface}
\acro{gpu}[GPU]{Graphics Processing Unit}
\acro{cpu}[CPU]{Central Processing Unit}
\acro{ir}[IR]{Intermediate Representation}
\acro{isa}[ISA]{Instruction Set Architecture}
\acro{jit}[JIT]{Just-in-Time}
\acro{llvm}[LLVM]{Low-Level Virtual Machine}
\acro{ptx}[PTX]{Parallel Thread Execution}
\acro{loc}[LOC]{lines of code}
\acro{nvrtc}[NVRTC]{NVIDIA Runtime Compilation}

\end{acronym}

%
%

\hypersetup{pdfauthor={Tim Besard},
            pdftitle={Effective Extensible Programming: Unleashing Julia on GPUs},
            pdfkeywords={Julia, CUDA}}

\definecolor{NavyBlue}{cmyk}{0.94,0.54,0,0}
\hypersetup{citecolor = NavyBlue,
            linkcolor = NavyBlue,
            urlcolor = NavyBlue}

\title{Effective Extensible Programming: \\ Unleashing Julia on GPUs}

\author{
  Tim Besard, Christophe Foket and Bjorn De Sutter, \IEEEmembership{Member,~IEEE}%
  \IEEEcompsocitemizethanks{\IEEEcompsocthanksitem T. Besard, C. Foket and
                            B. De Sutter are with the Department of
                            Electronics~and~Information~Systems,
                            Ghent~University, Belgium.\protect\\
  \texttt{\{tim.besard,christophe.foket\}@ugent.be}\protect\\
  Corresponding author: \texttt{bjorn.desutter@ugent.be}}
}

\markboth{IEEE Transactions on Parallel and Distributed Systems}%
{\ifCLASSOPTIONpeerreview Besard \MakeLowercase{\textit{et al.}}}

\newcommand\copyrighttext{%
  \centering \footnotesize This work has been submitted to the IEEE for possible
  publication. \\ Copyright may be transferred without notice, after which this
  version may no longer be accessible.}
\newcommand\copyrightnotice{%
\begin{tikzpicture}[remember picture,overlay]
  \node[anchor=south,yshift=10pt] at (current page.south) {
    \fbox{\parbox{\dimexpr\textwidth-\fboxsep-\fboxrule\relax}{\copyrighttext}}
  };
\end{tikzpicture}%
}



%

%

\IEEEtitleabstractindextext{%
\begin{abstract}

%
\acsp{gpu} and other accelerators are popular devices for accelerating
compute-intensive, parallelizable applications.
%
However, programming these devices is a difficult task. Writing efficient device
code is challenging, and is typically done in a low-level programming language.
High-level languages are rarely supported, or do not integrate with the rest of
the high-level language ecosystem.
%
To overcome this, we propose compiler infrastructure to efficiently add support
for new hardware or environments to an existing programming language.

We evaluate our approach by adding support for NVIDIA \acsp{gpu} to the Julia
programming language. By integrating with the existing compiler, we
significantly lower the cost to implement and maintain the new compiler, and
facilitate reuse of existing application code. Moreover, use of the
high-level Julia programming language enables new and dynamic approaches for
\acs{gpu} programming. This greatly improves programmer productivity, while
maintaining application performance similar to that of the official NVIDIA
\acs{cuda} toolkit.
%

\end{abstract}

\acresetall

}

\maketitle
\copyrightnotice

\IEEEpeerreviewmaketitle


%
%
\IEEEraisesectionheading{\section{Introduction}%
\label{sec:introduction}}
%
%


To satisfy ever higher computational demands, hardware vendors and software
developers look at accelerators, specialized processors that are optimized for
specific, typically parallel workloads, and perform much better at them than
general-purpose processors~\cite{kachris2016survey,
weerasinghe2015enabling,sarkar2010hardware, pratx2011gpu, aluru2014review}.
Multiple hardware vendors are working on such accelerators and release many new
devices every year. These rapid developments make it difficult for developers to
keep up and gain sufficient experience programming the devices. This is
exacerbated by the fact that many vendors only provide low-level toolchains,
such as \acs{cuda} or OpenCL, which offer full control to reach peak performance
at the cost of developer productivity~\cite{li2016comparing}.

To improve developer productivity, programmers commonly use high-level
programming languages. However, these languages often rely on techniques and
functionality that are hard to implement or even incompatible with execution on
typical accelerators, such as interpretation, tracing \ac{jit} compilation, or
reliance on a managed runtime library. To remedy this, implementations of
high-level languages for accelerators generally target a derived version of the
language, such as a restricted subset or an embedded \ac{dsl}, in which
incompatible features have been redefined or adjusted.

Modern extensible languages offer the means to realize such programming language
derivatives~\cite{zingaro2007modern}. For example, Lisp-like languages feature
powerful macros for processing syntax, Python's decorators make it possible to
change the behavior of functions and methods, and the Julia programming language
supports introspection of each of its \acp{ir}. However, these facilities do not
encourage reuse of existing compiler functionality. Most derived languages use a
custom compiler, which simplifies the implementation but hinders long-term
maintainability when the host language gains features or changes semantics. It
also forces users to learn and deal with the inevitable divergence between
individual language implementations.

This paper presents a vision in which the high-level language compiler exposes
interfaces to alter the compilation process (\Cref{sec:vision}). Implementations
of the language for other platforms can use these interfaces together with other
extensible programming patterns to ensure that source code is compiled to
compatible and efficient accelerator machine code. To demonstrate the power of
this approach, we have added such interfaces to the reference compiler of the
Julia language (\Cref{sec:interfaces}), and used it to add support for NVIDIA
\acsp{gpu} (\Cref{sec:cuda}). We show that the resulting toolchain makes it
possible to write generic and high-level \acs{gpu} code, while performing
similar to low-level \acs{cuda}~C (\Cref{sec:evaluation}). All code implementing
this framework is available as open-source software on GitHub, and can be easily
installed using the Julia package manager. Our contributions are as follows:

\begin{itemize}
    \item We introduce interfaces for altering the compilation process, and
    implement them in the Julia compiler.
    \item We present an implementation of the Julia language for NVIDIA
    \acsp{gpu}, using the introduced interfaces.
    \item We analyze the performance of \acs{cuda} benchmarks from the Rodinia
    benchmark suite, ported to Julia. We show that code generated by our
    \acs{gpu} compiler performs similar to \acs{cuda}~C code compiled with
    NVIDIA's reference compiler.
    \item We demonstrate high-level programming with this toolchain, and show
    that Julia \acs{gpu} code can be highly generic and flexible, without
    sacrificing performance.
\end{itemize}

\newpage

\section{Vision}
\label{sec:vision}
%
%



Our proposed solution to the difficulty in integrating high-level languages and
accelerators is a set of interfaces to the high-level language's general purpose
compiler, that provide fine-grained access to the different \acp{ir} and to the
processes that generate and optimize those \acp{ir}. With these interfaces,
developers can influence the existing language implementation and, e.g., improve
compatibility with new hardware or run-time environments without the need for a
custom compiler or an embedded language subset.

\begin{figure}[!t]
  \centering
  \includegraphics[scale=0.5]{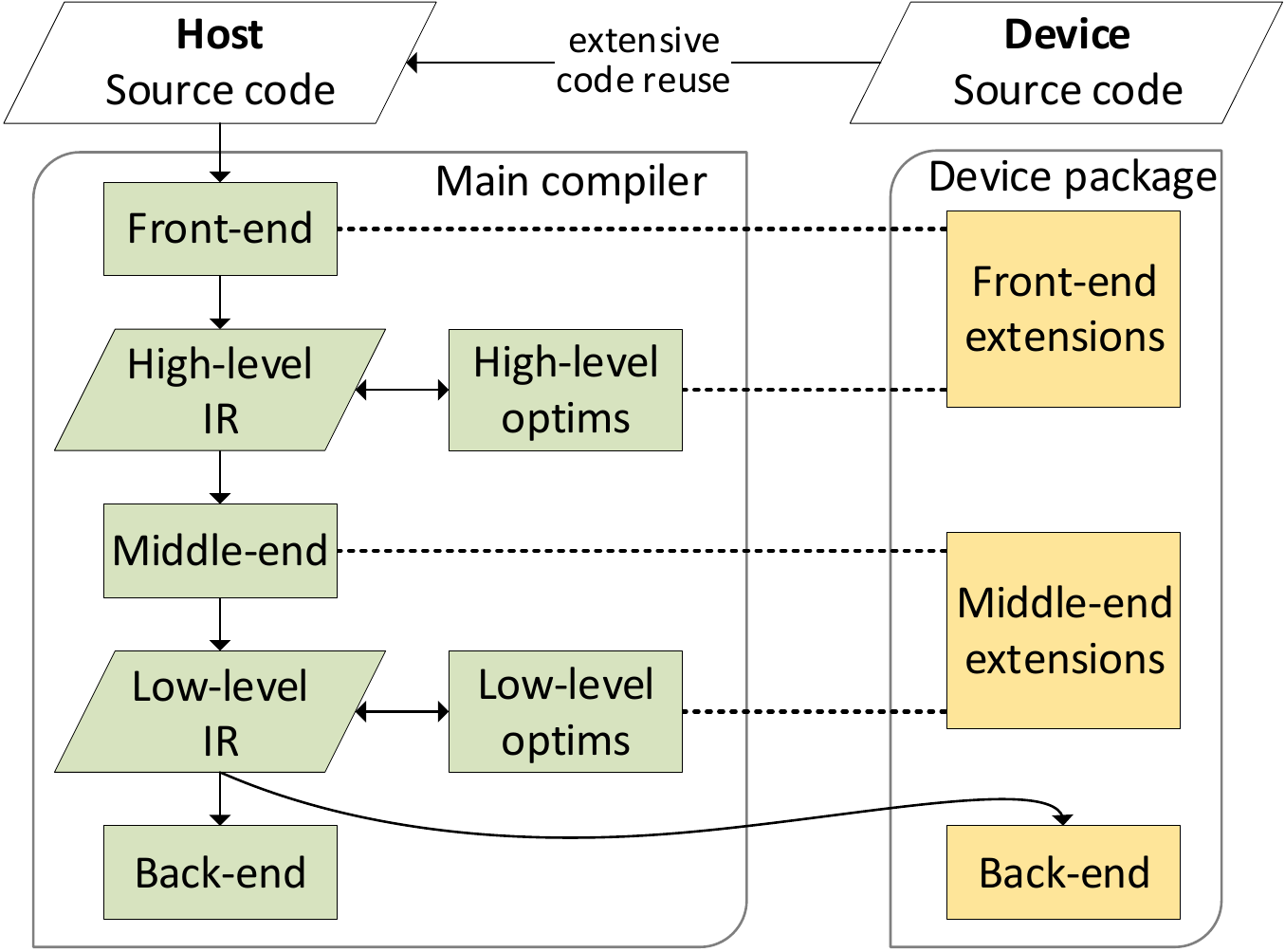}
  \vspace{-.5em}
  \caption{Abstract overview of the proposed toolchain.}
  \label{img:langimpl:abstract}
\end{figure}

\Cref{img:langimpl:abstract} shows an overview of the proposed toolchain. An
external device package uses the introduced interfaces to add support for new
hardware, without modifying the existing language implementation. For example,
it could refuse to generate code for certain language features, such as
exceptions or dynamic memory allocations, or replace their code with compatible
or optimized alternatives.


Such a setup has multiple advantages. For one, it keeps the existing language
implementation stable, while new implementations can be developed independently
as external packages. This makes it easier to experiment, as these packages do
not need to meet the support, quality, or licensing requirements of the existing
implementation. It also makes it easier to cope with the rapid development pace
of accelerator hardware, providing the means for vendors to contribute more
effectively to the language ecosystem.

\begin{figure}[!t]
  \centering
  \includegraphics[scale=0.5]{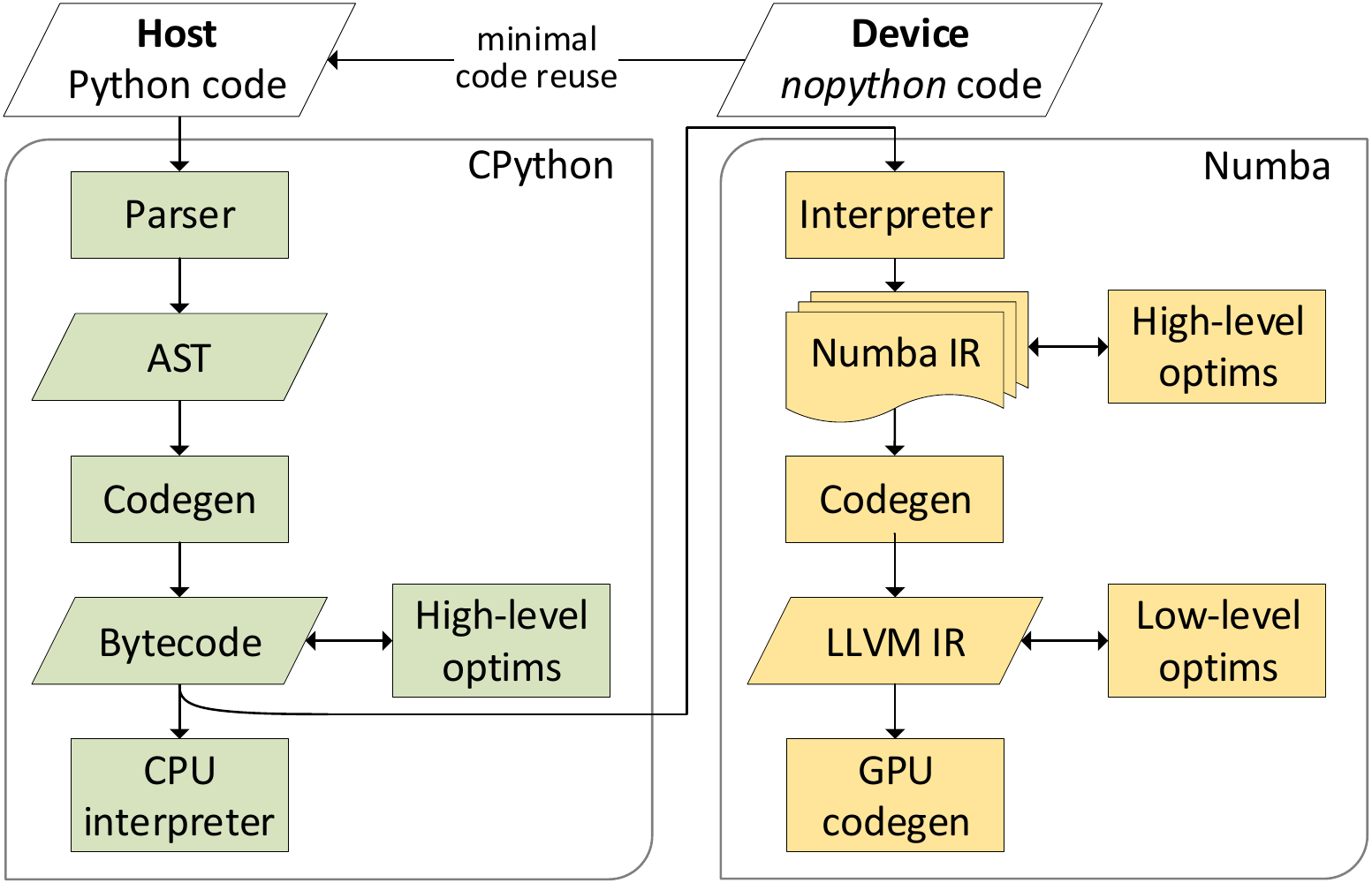}
  \vspace{-.5em}
  \caption{CPython and Numba compilation processes for host and device.}
  \label{img:langimpl:python}
\end{figure}

Another important advantage is the ability to reuse the existing language
implementation, whereas current implementations of high-level languages for
accelerators often reimplement large parts of the compiler. For example, Numba
is a \ac{jit} compiler for Python, building on the CPython reference language
implementation. As \Cref{img:langimpl:python} shows, the Numba compiler takes
Python bytecode and compiles it to optimized machine code. Due to the high-level
nature of Python bytecode, the Numba interpreter and subsequent compilation
stages duplicate much functionality from CPython: CFG construction, type
inference, liveness analysis, and implementations of built-in functions. As a
result, each release of Numba is tailored to the specifics of certain CPython
versions~\cite{lam2015numba}, and needs to be updated when changes are made to
the language implementation. The semantics of code also differ slightly
depending on whether it is interpreted by CPython or compiled with
Numba~\cite{lam2015numba}, further impeding compatibility with existing Python
code.

Our proposed compiler interfaces make it possible to share functionality
between an existing language implementation and external derivatives, avoiding
needless reimplementation of compiler functionality by reconfiguring the
existing compiler to generate code that is compatible with the platform at hand.
This not only facilitates external language implementations, but also improves
compatibility with existing code as it avoids the inevitable differences between
individual compiler implementations.

In some cases, even more reuse of existing infrastructure than suggested in
\Cref{img:langimpl:abstract} is possible. When the back-end compiler used in the
general-purpose tool flow can also target accelerators, there is no need to
reimplement a device back end in the device package. Instead, that existing
back-end compiler can then be used for host and device code. Even if this is not
the case, it might not be necessary to reimplement a full device back end in
the device package: If third-party device code generators can be reused, the
device back end only has to translate the low-level IR code to an IR accepted by
that third-party code generator.


Conceptually, the compiler interfaces shown in \Cref{img:langimpl:abstract} are
generally applicable. Their actual instantiation, however, will be specific to
the host language and accelerator at hand. We expect further research into such
interfaces to generalize the design and improve reusability across languages and
accelerators. For now, we will design the interfaces around a single language
and accelerator platform.

For this work, we chose to target \acp{gpu}, massively parallel accelerators
with a distinctive enough architecture to make optimization worthwhile, yet
broadly usable for many kinds of applications (as explained in
\Cref{sec:background:gpu}). Specifically, we focus on NVIDIA \acp{gpu} with the
\ac{cuda}, because of the mature toolchain and hardware availability. We target
this hardware from the Julia programming language, a high-level technical
computing language built on the \ac{llvm} compiler framework. As we explain in
\Cref{sec:background:julia}, this language is a good fit for accelerator
programming, while offering flexible tools to extend the language, e.g., for the
purpose of targeting new hardware. Furthermore, given its use of \ac{llvm}, and
\ac{llvm}'s capabilities to target both CPUs and \ac{cuda} \acp{gpu}, we will
not need to reimplement a device back end ourselves, as mentioned above.

\section{Background}
\label{sec:background}
%
%


\subsection{GPU Accelerators}
\label{sec:background:gpu}

\acp{gpu} are massively parallel accelerators that can speed up
compute-intensive general-purpose applications. However, that generality is
constrained: Most \acp{gpu} need to be treated like a coprocessor (with separate
memory spaces, controlled by a host processor, mostly unable to perform
input/output operations, etc.), and can only efficiently execute codes that
exhibit specific kinds of parallelism. As a result, \acp{gpu} are relatively
hard to program: Programmers have to deal with the intricacies of coprocessor
programming, and need experience with parallel programming to assess if and how
specific problems can be solved effectively on a \ac{gpu}.

At the same time, vendor-supported development environments for programming
\ac{gpu} accelerators typically work with low-level programming languages.
NVIDIA's \ac{cuda}, for instance, uses \ac{cuda}~C, while AMD and Intel
\acp{gpu} are programmed using OpenCL~C. The constructs in these low-level
languages map closely to available hardware features, making it possible to
reach peak performance, as potentially costly abstractions are avoided. However,
the lack of such abstractions also complicates \ac{gpu} programming, not only
requiring parallel programming skills and domain knowledge to map the problems,
but also low-level programming competence and \ac{gpu} hardware know-how for the
actual implementations~\cite{li2016comparing}. Furthermore, due to a lack of
abstractions, these implementations are often hardware-specific, or perform
significantly worse on different hardware~\cite{du2012cuda}. Libraries like
CUB~\cite{merrill2015cub} or Thrust~\cite{hoberock2010thrust} aim to raise the
abstraction level and portability using C++~templates, but fall short due to the
low-level nature of C++ and limited applicability across vendor toolkits.

Rather than programming accelerators directly, developers can also use optimized
host libraries that are called from the host processor and not directly from the
device. Hardware vendors provide such libraries, implementing popular interfaces
like BLAS~\cite{dongarra1990set} and LAPACK~\cite{anderson1999lapack}. There
also exist third-party libraries like ArrayFire~\cite{malcolm2012arrayfire} and
ViennaCL~\cite{rupp2010viennacl} that abstract over devices and platforms. These
libraries typically export a C~\ac{api}, which eases their use outside of the
vendor-supplied development environment. For example, the \ac{cuda} BLAS library
cuBLAS~\cite{nvidia2008cublas} can be used from
Python~\cite{continuum2017accelerate}, Julia~\cite{julia2017cublas},
Octave~\cite{nvidia2015octave}, etc. However, compilers for these languages
cannot reason about code in the libraries, and they cannot optimize code across
calls to it. Moreover, library-driven development requires programming in terms
of abstractions, which are typically coarse-grained to amortize the cost of
configuring the accelerator, initiating execution, etc. Most libraries are also
unable to compose their abstractions with custom device code. As a result,
library-based programming can be unfit for implementing certain types of
applications.

Using high-level languages to program accelerators directly provides a middle
ground between high-level host libraries and direct programming with vendor
toolkits: Direct programming can offer fine-grained control over compilation and
execution, while the use of a high-level language and its abstraction
capabilities can improve programmer productivity. However, existing
implementations of high-level languages for accelerators do not integrate well
with the rest of the language. Some come in the form of an embedded \ac{dsl},
such as PyGPU or Copperhead~\cite{lejdfors2007pygpu, catanzaro2011copperhead},
which programmers have to learn and to which they have to adapt their code.
Continuum Analytics' Numba~\cite{lam2015numba} reimplements support for a subset
of the Python language that is appropriately called \code{nopython} because it
does not support many of the high-level features of Python because these
features do not map well onto \acp{gpu}, while duplicating compiler
functionality from the CPython reference implementation as shown in
\Cref{img:langimpl:python}. Our proposed interfaces serve to avoid this
duplication, and integrate with the existing language implementation for the
purpose of improved code compatibility and more effective compiler
implementation.

\subsection{Julia Programming Language}
\label{sec:background:julia}

Julia is a high-level, high-performance dynamic programming language for
technical computing~\cite{bezanson2017julia}. It features a type system with
parametric polymorphism, multiple dispatch, metaprogramming capabilities, and
other high-level features~\cite{bezanson2015abstraction}. The most remarkable
aspect of the language and its main implementation is speed: carefully written
Julia code performs exceptionally well on traditional microprocessors,
approaching the speed of code written in statically-compiled languages like C or
FORTRAN~\cite{bezanson2012julia, besard2015case}.



Julia's competitive performance originates from clever language design that
avoids the typical compilation and execution uncertainties associated with
dynamic languages~\cite{bezanson2015fast}. For example, Julia features a
systemic vocabulary of types, with primitive types (integers, floats) mapping
onto machine-native representations. The compiler uses type inference to
propagate type information throughout the program, tagging locations (variables,
temporaries) with the type known at compile time. If a location is fully typed
and the layout of that type is known, the compiler can often use stack memory to
store its value. In contrast, uncertainty with respect to the type of a location
obligates variably-sized run-time heap allocations, with type tags next to
values and dynamic checks on those tags as is common in many high-level
languages.

\begin{span}
  \begin{lstlisting}[language=julia, label={lst:julia:branches},
                     caption={Single-dispatch polymorphism and branches that
                              leads to unstable functions, returning
                              differently-typed objects based on run-time
                              values.},
                     gobble=4]
    function intersect(a::Rect, b)  # returns Rect or Line
      if isa(b,Rect)
        return c::Rect
      else if isa(b,Line)
        return c::Line
      end
    end
    function intersect(a::Line, b)  # returns Rect or Line
      return c
    end
  \end{lstlisting}
  \begin{lstlisting}[language=julia, label={lst:julia:multimethods},
                     caption={Functionality of \Cref{lst:julia:branches}
                              expressed through multiple dispatch, with
                              narrowly-typed methods.},
                     gobble=4]
    function intersect(a::Rect, b::Rect)    # returns Rect
      return c::Rect
    end
    function intersect(a::Rect, b::Line)    # returns Line
      return c::Line
    end
  \end{lstlisting}
\end{span}

Similarly, types are used to express program behavior and eliminate execution
uncertainty by means of multiple dispatch or multimethods. This type of function
dispatch selects an appropriate method based on the run-time type of all of its
arguments, and is a generalization of single-dispatch polymorphism (e.g., as
seen in C++) where only the object on which a method is called is used to
disambiguate a function call. For example, \Cref{lst:julia:branches} does not
use multiple dispatch and defines \code{intersect} methods that only dispatch on
the first argument, returning differently-typed objects by branching on the type
of values. Conversely, \Cref{lst:julia:multimethods} defines multiple methods
that dispatch on all arguments, and consequently are more narrowly-typed in
terms of arguments and returned values. In the case of a sufficiently typed
call, this enables the compiler to dispatch statically to the correct method and
avoid run-time branches, possibly even stack-allocating the returned value if
its layout is known.

\begin{figure}[!t]
  \centering
  \includegraphics[scale=0.5]{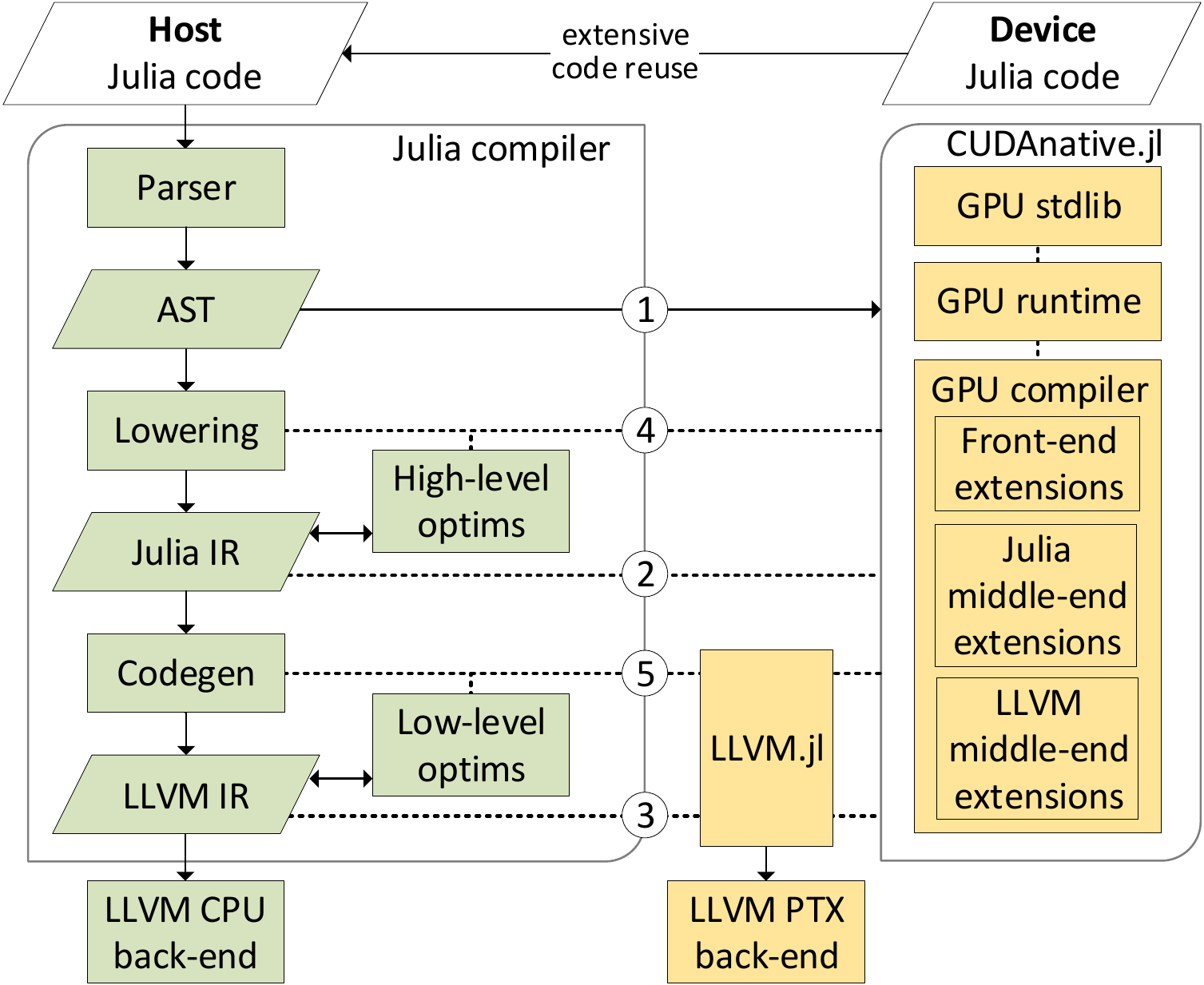}
  \vspace{-.5em}
  \caption{Schematic overview of the compilation process for Julia \acs{gpu}
           code with CUDAnative.jl, building on the existing compiler by means
           of extension interfaces. Dashed arrows indicate generic interactions,
           while solid arrows represent the flow of code.}
  \label{img:langimpl:julia}
  \vspace{-.5em}
\end{figure}

The combination of this design and aggressive specialization on run-time types
enables the Julia compiler to generate mostly statically-typed intermediate
code, without the need for \ac{jit} compilation techniques traditionally used by
high-level language implementations (tracing, speculative execution,
deoptimization, etc.). This allows the Julia developers to outsource the
back-end part of the compilation flow to existing compiler frameworks for static
languages. In particular, the Julia \ac{ir} is a good fit for the \ac{llvm}
compiler framework, which is commonly used as a basis for industrial-strength
compilers for static languages~\cite{lattner2004llvm}. The Julia compiler
targets this framework by emitting \ac{llvm}~\ac{ir} as the low-level \ac{ir}
from \Cref{img:langimpl:abstract}, and uses the vast array of \ac{llvm}
optimization passes (often tailored for or assuming statically-typed
straight-line \ac{ir}) to optimize code and ultimately compile it to
high-performance CPU machine code. The left part of \Cref{img:langimpl:julia}
shows this existing Julia compilation tool flow. In the remainder of this paper,
we refer to it as the main compiler because it is the part of the flow that will
generate machine code for the main, general-purpose CPU(s) that serve as a host
to accelerators. The last main processing step, CPU code generation, is
implemented entirely by means of \ac{llvm}. To facilitate interactions with this
C++ library, those parts of the Julia compiler that interface with \ac{llvm} are
also written in C++, making it possible to directly invoke its \acp{api}.

As a testament to the performance this design can achieve, most of the Julia
standard library is written in Julia itself (with some obvious exceptions for
the purpose of reusing existing libraries), while offering good
performance~\cite{bezanson2012julia, besard2015case}. The managed runtime
library is only required for dynamic code that might trigger compilation, and
certain language features such as garbage collection and stack unwinding.

Coincidentally, this design also makes the language well-suited for accelerator
programming. Such hardware often features a different architecture and \ac{isa},
operating independently from the main processor, with control and data transfers
happening over a shared bus. In many cases, this makes it hard or impossible to
share code, such as runtime libraries, between host and device. With Julia,
however, it is entirely possible to write high-level code that boils down to
self-contained, static IR, a prerequisite for many accelerator programming
models.

\begin{table}[!t]
  \renewcommand{\arraystretch}{1.2}
  \caption[]{Existing metaprogramming interfaces in Julia
             to access compiler \acp{ir}.}
  \label{tbl:interfaces:ir}
  \centering
  \vspace{-1em}
  \begin{tabular}{|c|c c|}
    \hline
                                     & Access    & Modify                     \\
    \hline
    \circled{1} \acs{ast}            & \cmark    & \cmark                     \\
    \circled{2} Julia \acs{ir}       & \cmark    & \cmark                     \\
    \circled{3} \acs{llvm} \acs{ir}  & \cmark    & \cmark                     \\
    Machine code                     & \cmark    & indirectly                 \\
    \hline
  \end{tabular}
  \vspace{-1em}
\end{table}

In addition, Julia features powerful metaprogramming and reflection
capabilities, as shown in \Cref{tbl:interfaces:ir}. Source code can be
introspected and modified using \emph{macros}, or using the \code{parse} and
\code{eval} functions. The high-level Julia \ac{ir} is accessible with the
\code{code_lowered} and \code{code_typed} reflection functions, and can be
modified with \emph{generated functions}. These mechanisms are powerful,
flexible, and user-friendly, because they have been co-designed together with
the source language and the tool flow in support of metaprogramming and
reflection, and because Julia is a homoiconic programming language, i.e., code
can be accessed and transformed as structured data from within the language. As
such, these interfaces already offer some of the flexibility required to target
new hardware, e.g., to define constructs with non-standard semantics or special
code generation without the need for new language features. As we will discuss
in the next section, however, their support does not yet suffice for targeting
accelerators like GPUs.

Low-level \ac{llvm}~\ac{ir} can be inspected by invoking \code{code_llvm} and
injected via the \code{llvmcall} metaprogramming interface. Machine code is
accessible through \code{code_native} and can be inserted indirectly as inline
assembly in \ac{llvm}~\ac{ir}. These interfaces are much less powerful and
flexible, however. Most importantly, the interfaces to \ac{llvm}~\ac{ir} only
pass string representations of the \ac{ir} code. This generic and neutral form
of interface fits the separation of concerns between Julia and \ac{llvm}. It
suffices for the main compiler because (i) metaprogramming and reflection do not
require interfaces at the \ac{llvm}~\ac{ir} level, (ii) \code{llvmcall} is
currently only used to inject small, literal snippets of \ac{llvm}~\ac{ir},
e.g., to add support for atomics, and (iii) the main compiler is implemented
mostly in C++, and thus has direct access to the \ac{llvm}~\ac{ir} builder
interfaces.

However, as we will discuss in the next section, these string-based interfaces
to the lower-level \acp{ir} do not suffice for targeting accelerators from
within device packages.

\section{Effective Extensible Programming}
\label{sec:interfaces}

As discussed in \Cref{sec:vision}, we propose to integrate high-level
programming languages with accelerator hardware by means of extension interfaces
to the existing compiler that was, and will continue to be, developed by and
large to target general-purpose hardware. The existing interfaces to manipulate
the different \acp{ir} as discussed in the previous section provide a good
starting point, but they do not yet suffice.

First, although they make it possible to improve compatibility with accelerators
by manipulating general purpose \ac{ir} or generating compatible \ac{ir} from
scratch, they fall short in reusing and repurposing the main compiler's
\ac{ir}-generating components. \Cref{sec:interfaces:front-end} proposes our
solution to make the compiler generate accelerator-optimized \ac{ir} in the
first place.

Secondly, the string-based interfaces to the lower-level \acp{ir} do not let the
device package reuse main compiler functionality to generate and inject
low-level \ac{ir} code. As targeting accelerators such as GPUs requires more
than injecting predetermined code snippets, this lack of reuse is problematic.
\Cref{sec:interfaces:back-end} presents a solution to this issue.

\subsection{Front-end IR Interfaces}
\label{sec:interfaces:front-end}

As a starting point to generate accelerator code, the language should offer
access to the different representations of source code in the compiler, such as
syntax trees, \acp{ir}, and machine code. This makes it possible to implement
functionality that cannot be expressed in the source language by manually
emitting intermediate code, without the need to alter the language or compiler.
It can also be used to transform \ac{ir}, or use it as a starting point for
further compilation. In the case of the Julia programming language, there
already exist several metaprogramming interfaces that provide access to those
intermediate forms of code, as shown in \Cref{tbl:interfaces:ir}.

For the purpose of external language implementations, simply having access to
the code generated for each \ac{ir} level is insufficient. In addition, access
to the \ac{ir} code should be augmented with access to the very processes that
generate that code. For example, when compiling code for an environment that
does not support the Julia runtime library, the compiler needs to avoid calls to
it. A typical case is that of exceptions, which rely on the runtime for stack
unwinding and error reporting.  In the main Julia compiler, these calls to the
runtime are generated as part of the code generation process that lowers
Julia~\ac{ir} to \ac{llvm}~\ac{ir}. To generate code that does not require the
runtime library without altering the code generation process, the compiler needs
to rid the Julia~\ac{ir} from exceptions, or remove calls to the runtime from
the generated \ac{llvm}~\ac{ir}. Both approaches are fragile, because they
involve modeling behavior of the main compiler and duplicating functionality
from it.

To overcome this problem and improve the reusability of the compiler, we added
the four interfaces from \Cref{tbl:interfaces:irgen} that offer additional
control over code generation processes. More specifically, both the lowering of
\acp{ast} to Julia~\ac{ir}, and Julia~\ac{ir} to \ac{llvm}~\ac{ir} can now be
altered through parameters and hooks to reconfigure or replace individual
components of these code generation processes. Applied to the above example of
code generation without a runtime library, a so-called \code{CodegenParam} could
be used to disallow exceptions altogether, or alternatively a \code{CodegenHook}
could change the generated code not to rely on the runtime library. The \ac{gpu}
back end from \Cref{sec:cuda} uses these interfaces to replace or customize code
generation functionality for exceptions, dynamic memory allocation such as
garbage collection, and other functionality that typically requires runtime
support libraries, Of course, the nature of these parameters and hooks are
specific to the language and its compiler, but the approach is generic and
enables extensive reuse of existing functionality.

\begin{table}[!t]
  \renewcommand{\arraystretch}{1.2}
  \caption[]{Additional interfaces to the Julia compiler \\
             for controlling code generation processes.}
  \label{tbl:interfaces:irgen}
  \centering
  \vspace{-0.5em}
  \begin{tabular}{|c|c c|}
    \hline
                                    & Reconfigure              & Replace                 \\
    \hline
    \acs{ast}                       & -                        & -                       \\
    \circled{4} Julia \acs{ir}      & \texttt{InferenceParams} & \texttt{InferenceHooks} \\
    \circled{5} \acs{llvm} \acs{ir} & \texttt{CodegenParams}   & \texttt{CodegenHooks}   \\
    Machine code                    & -                        & -                       \\
    \hline
  \end{tabular}
\end{table}

For now, we have only introduced such interfaces to the processes that generate
Julia and \ac{llvm}~\ac{ir}; The parsing phase that converts source-code to an
\ac{ast} is superficial and generic enough not to need adjustment for \ac{gpu}
execution, while machine code generation is extremely target-specific and does
not offer many opportunities for reuse.

\subsection{Back-end IR Interfaces}
\label{sec:interfaces:back-end}

The codegen step in the main compiler translates (i.e., lowers) Julia \ac{ir}
constructs into \ac{llvm}~\ac{ir}. The C++ part of the codegen implementation
directly invokes \ac{llvm}~\ac{ir} builder interfaces to do so; the part
implemented in Julia itself uses the aforementioned string-based interfaces.

For the device package in support of an accelerator target, we want to avoid
both mechanisms as much as possible. The string-based approach is too fragile,
and the C++ approach is not productive enough for the developer of the device
package. This developer can be expected to be an expert in Julia and in his
targeted accelerators, but not in C++ programming or C++ \acp{api}. For this
reason, we strive for providing the necessary interfaces and functionality that
lets developers create new language implementations (for accelerators) in the
Julia language itself, and that shields them from as many \ac{llvm} details as
possible. This greatly lowers the required effort to support new hardware, as
much less code is required when the necessary accelerator-oriented compiler
functionality can be written in a productive programming language. As a
testament to the value of this goal, the \ac{gpu} support presented in
\Cref{sec:cuda} only requires about 1500 \ac{loc}.

Furthermore, no changes to the language's compiler are then required, which
enables the distribution of the new language implementations (i.e., the device
packages) independent from the existing implementation, e.g., with a built-in
package manager. The new implementation can be quickly iterated and improved
upon, while keeping the core language and its compiler stable. Such a
development model is especially interesting for accelerator vendors, where the
rapid pace of hardware developments necessitates frequent changes to the
toolchain. This contrasts with the relatively slow developments in host
compilers and with sometimes conservative upgrade policies by system
administrators.

To facilitate interactions with \ac{llvm}, we have created the \emph{LLVM.jl}
package, which is available at \url{https://github.com/maleadt/LLVM.jl}. It
provides a high-level wrapper to the \ac{llvm} C \ac{api}, using Julia's
powerful \ac{ffi} to interact efficiently with the underlying libraries. The
package can be used to inspect, modify or emit \ac{llvm}~\ac{ir}. It greatly
improves the usability of the extension interfaces that operate at the
\ac{llvm}~\ac{ir} level. In addition, the package enables reuse of back-end
compiler functionality that are part of \ac{llvm}, including the vast array of
optimization passes that are part of \ac{llvm}, or the many back ends to
generate machine code from \ac{llvm}~\ac{ir}.

As an example of using LLVM.jl for the purpose of implementing the necessary
lowering from Julia IR to \ac{llvm}~\ac{ir}, \Cref{lst:llvm:generated} shows how
to implement a custom function for loading values from a pointer using the
LLVM.jl interfaces. In Julia IR, accessing, e.g., an element in an array, is
modeled with a call to the \code{unsafe_load} function. This function is
implemented in the standard Julia library. Its body contains a call to an
intrinsic function that is recognized by the codegen processing step in the main
Julia compiler, which then lowers it to appropriate \ac{llvm}~\ac{ir} code.

Implementing an optimized version of \code{unsafe_load} for loading values on
accelerators using the same intrinsics mechanism would similarly require the
introduction of one or more intrinsics in the main compiler, and writing the
necessary lowering support in C++ using \ac{llvm}~\acp{api}. This is cumbersome,
inflexible, and unproductive.

By contrast, the code in \Cref{lst:llvm:generated} shows how to load a value
from a pointer with Julia metaprogramming and the LLVM.jl package.\footnote{To
avoid uninteresting clutter in our illustration of LLVM.jl, \\ we show a simplified
\code{load} function instead of the full \code{unsafe_load}.}  It is implemented
using a generator function, declared with \code{@generated} on
\cref{lst:llvm:generated:fun}, which builds the expressions that should be
executed at run time. Generator functions are invoked during type-inference, for
every combination of argument types the function is invoked with. In this case,
this function generates \ac{llvm}~\ac{ir} and returns an \code{llvmcall}
expression that injects the code into the compiler, effectively returning the
\ac{ir} that will have to be executed in the application. Note that this is much
stronger than using macros: on \cref{lst:llvm:generated:type}, the pointer
argument \code{p} is not only known by name, but its type \code|Ptr{T}| as
determined by type inference in the Julia compiler is also known to the
generator function, with \code{T} being a type variable referring to the actual
runtime element type of the pointer. The generated code hence depends on the
inferred types, and can be customized and optimized for it at each invocation of
\code{load} in the Julia IR. In the next section, we will discuss how this can
be exploited to generate memory accesses optimized for the different types of
memories in a \ac{gpu} memory hierarchy.

Without the LLVM.jl interface, the load function body would have been full of
string manipulations, which would have been a nightmare in terms of code
readability. Moreover, it would have contained cases for every supported pointer
type, and the optimization for, e.g., different types of memories, would be hard
or impossible.

\begin{lstlisting}[float, language=julia, label={lst:llvm:generated},
                   caption={Using LLVM.jl to implement functions that generate
                            their own \acs{ir}, specialized on the types of the
                            arguments.},
                   gobble=2]
  # loading a value from a pointer
  @generated function load(p::Ptr{T}) where {T}¶\label{lst:llvm:generated:fun}¶
    eltyp = LLVM.convert(LLVM.Type, T)¶\label{lst:llvm:generated:type}¶
 
    # create a LLVM module and function
    mod = LLVM.Module("llvmcall")
    param_typs = [LLVM.PointerType(eltyp)]
    ft = LLVM.FunctionType(eltyp, param_typs)
    f = LLVM.Function(mod, "load", ft)
 
    # generate IR
    LLVM.Builder() do builder
      bb = LLVM.BasicBlock(f, "entry")
      LLVM.position!(builder, bb)
 
      ptr = LLVM.parameters(f)[1]
      val = LLVM.load!(builder, ptr) # the actual load
 
      LLVM.ret!(builder, val)
    end
 
    # inject the IR and call it
    return :( llvmcall($f, $T, Tuple{Ptr{$T}}, p) )
  end
 
  @test load(pointer([42])) == 42
\end{lstlisting}


%
%
\section{CUDA Language Implementation}
\label{sec:cuda}

To eat our own dog food, we used the infrastructure from \Cref{sec:interfaces}
to developed a \ac{gpu} implementation of the Julia programming language that
targets NVIDIA hardware via the \ac{cuda} toolkit. This implementation is an
instantiation of the device package shown in \Cref{img:langimpl:abstract}. It is
distributed as a regular Julia package named \emph{CUDAnative.jl}, which is
available at \url{https://github.com/JuliaGPU/CUDAnative.jl}, and does not
require any modifications to the underlying Julia compiler. It supports a subset
of the Julia language, but that subset has proven extensive enough to implement
real-life \ac{gpu} applications and build high-level abstractions.

The device package actually consists of three major components, as shown on the
right of \Cref{img:langimpl:julia}: a standard library of \ac{gpu}-specific
functionality, a compiler to generate \ac{gpu} machine code from Julia sources,
and a runtime system to invoke the compiler and manage it together with the
underlying \ac{gpu} hardware. Together with the main compiler, which serves as a
JIT compiler for host CPUs, this package serves as a JIT compiler for
\ac{cuda}~\acp{gpu}.

\subsection{Standard Library}
\label{sec:cuda:stdlib}

The CUDAnative.jl standard library focuses on providing definitions for
low-level \ac{gpu} operations that are required for writing effective \ac{gpu}
applications. For example, to access registers containing current thread and
block indexes, define synchronization barriers, or allocate shared memory.

Whereas many languages would implement these definitions using compiler
intrinsics -- built-in functions whose implementation is handled specially by
the compiler -- the Julia programming language is expressive enough to implement
much of this functionality using Julia code itself. Built-in functions might
still be necessary to implement very low-level interactions, but the amount of
these functions and their responsibilities are greatly reduced. For example,
where CPython implements the \code{print} function entirely in C as part of the
compiler, Julia only relies on a \code{write} function to write bytes to
standard output.

Even when the language isn't expressive enough, intrinsics can be avoided by
generating lower-level code directly using the metaprogramming interfaces from
\Cref{tbl:interfaces:ir}. For example, atomics are implemented with literal
snippets of \ac{llvm}~\ac{ir} and wrapped in user-friendly language constructs
by means of macros. The \ac{gpu} standard library in CUDAnative.jl relies
heavily on this type of programming, with help from the \ac{llvm} \ac{api}
wrapper from \Cref{sec:interfaces:back-end} to facilitate interactions with the
\ac{llvm}~\ac{ir}.

Julia's expressiveness and metaprogramming functionality make it possible for
most of the Julia standard library to be written in Julia itself. This makes the
standard library much easier to extend or override, e.g., using type-based
multiple dispatch as demonstrated in \Cref{lst:julia:multimethods}.
CUDAnative.jl relies on this extensibility to improve compatibility or
performance of existing language features, as the next section illustrates.

\subsubsection{Pointers with Address Spaces}
\label{sec:cuda:stdlib:devptr}

Pointer address spaces identify, in an abstract way, where pointed-to objects
reside. They may be used for optimization purposes such as identifying pointers
to garbage-collected memory, or may have a physical meaning depending on the
hardware being targeted. In the case of \ac{ptx} code emitted for NVIDIA
\acp{gpu}, address spaces are used to differentiate between state spaces:
storage areas with particular characteristics in terms of size, access speed,
sharing between threads, etc. The \ac{ptx} compiler will use this information to
emit specialized memory operations, such as \code{ld.global} or
\code{st.shared}.  If no address space is specified, untagged operations will be
emitted (\code{ld} or \code{st}) which make the \ac{gpu} determine the state
space at run time by checking against a memory window. While implementing
initial \ac{cuda} support for Julia, we observed that these untagged operations
significantly lower the performance of memory-intensive benchmarks.

\ac{llvm} already includes optimizations to infer address space information
across memory operations~\cite{wu2016gpucc}, but these fall short when the
memory allocation site is invisible. For example, pointers can be passed as
arguments to a kernel, in which case the allocation happened on the host and is
invisible to the \ac{gpu} compiler. This is very common with \ac{gpu} codes,
where entry-point kernels often take several (pointers to) arrays as arguments.

\begin{lstlisting}[float, language=julia, label={lst:cudanative:pointer},
                   caption={Implementation of optimized \acs{gpu} pointers in
                            CUDAnative.jl, building on the example from
                            \Cref{lst:llvm:generated}.},
                   gobble=2]
  # custom pointer with address-space information
  struct DevPtr{T,AS}¶\label{lst:cudanative:pointer:struct}¶
    ptr::Ptr{T}
  end

  # loading an indexed value from a pointer
  @generated function unsafe_load(p::DevPtr{T,AS},¶\newline¶                                i::Int=1)¶\newline¶                    where {T,AS}¶\label{lst:cudanative:pointer:fun}¶
    eltyp = LLVM.convert(LLVM.Type, T)

    # create a LLVM module and function
    ...

    # generate IR
    LLVM.Builder() do builder
      ...
      # load from ptr with AS
      ptr = LLVM.gep!(builder, LLVM.parameters(f)[1],¶\newline¶                    [parameters(f)[2]])
      devptr_typ = LLVM.PointerType(eltyp, AS)
      devptr = LLVM.addrspacecast!(builder, ptr,¶\newline¶                                 devptr_typ)¶\label{lst:cudanative:pointer:cast}¶
      val = LLVM.load!(builder, devptr)¶\label{lst:cudanative:pointer:load}¶
      ...
    end

    # inject the IR and call it
    ...
  end
\end{lstlisting}

In Julia, pointers are represented by \code{Ptr} objects: regular objects with
no special meaning, and operations on these pointers are implemented using
normal methods. As such, we can easily define our own pointer type.
\Cref{lst:cudanative:pointer} shows how CUDAnative.jl provides a custom
\code{DevPtr} type representing a pointer with address-space information. By
implementing the excepted method interface, which includes the
\code{unsafe_load} method defined on \cref{lst:cudanative:pointer:fun},
\code{DevPtr} objects can be used in place of \code{Ptr} objects. This then
yields specialized memory operations that perform better.

The implementation of \code{unsafe_load} in \Cref{lst:cudanative:pointer} uses
the metaprogramming techniques explained in \Cref{sec:interfaces:back-end}. A
generator function builds specialized \ac{llvm}~\ac{ir} and injects it back in
the compiler, with the relevant address-space-specific load on
\cref{lst:cudanative:pointer:cast, lst:cudanative:pointer:load}. This allows to
implement low-level functionality that cannot be expressed using pure Julia
code, without the need for additional compiler intrinsics.

Note how the \code{DevPtr} type from \cref{lst:cudanative:pointer:struct} only
contains a single \code{ptr} field and as such has the exact same memory layout
as the existing \code{Ptr} type. The address space information is only known by
the type system, and does not affect the memory representation of run-time
pointers.

\subsubsection{NVIDIA Device Library}
\label{sec:cuda:stdlib:libdevice}

Another important source of low-level \ac{gpu} operations in CUDAnative.jl is
\code{libdevice}, a bitcode library shipped as part of the \ac{cuda} toolkit.
This library contains common functionality implemented for NVIDIA \acp{gpu},
including math primitives, certain special functions, bit manipulation
operations, etc. The CUDAnative.jl package provides wrappers for these
operations, compatible with counterpart functionality in the Julia standard
library. This often raises the abstraction level, and improves usability. For
example, \code{libdevice} provides 4 different functions to compute the absolute
value: \code{__nv_abs} and \code{__nv_llabs} for respectively 32-bit and 64-bit
integers, and similarly \code{__nv_fabs} and \code{__nv_fabsf} for 32-bit and
64-bit floating-point values. The Julia wrapper provides the same functionality,
but as different methods of a single generic function \code{abs}.

\subsection{GPU Compiler}
\label{sec:cuda:compiler}




Together with the main Julia compiler, the CUDAnative.jl infrastructure of
\Cref{img:langimpl:julia} instantiates the design from
\Cref{img:langimpl:abstract}, with respectively the Julia \ac{ir} and
\ac{llvm}~\ac{ir} as the high and low-level \acp{ir}. Together with host Julia
code, device code is processed by the main compiler's parser, which lowers
syntactical constructs and expands macros. Both the host code and the device
code can include application code as well as library code, and there is no
inherent difference between either type of code. There is no need for an
explicit annotation or encapsulation of device code, greatly improving
opportunities for code reuse. For example, barring use of incompatible language
features, much of the Julia standard library can be used to implement device
code.

The main interface for calling functions on a \ac{gpu} resembles a call to an
ordinary Julia function: \code{@cuda (config...) function(args...)}, where the
\code{config} tuple indicates the launch configuration similar to the triple
angle bracket syntax in \acs{cuda}~C. Because of the way \code{@cuda} is
implemented in the \ac{gpu} standard library using metaprogramming, the Julia
compiler invokes the \ac{gpu} compiler in CUDAnative.jl whenever such a call
occurs in the code. That \ac{gpu} compiler then takes over the compilation of
the called code. Using the existing interfaces from \Cref{tbl:interfaces:ir},
the new interfaces from \Cref{tbl:interfaces:irgen}, and the LLVM.jl wrapper,
the \ac{gpu} compiler configures and invokes the existing main compiler
components for lowering the (expanded) \ac{ast} into \ac{gpu}-oriented Julia IR,
for performing high-level optimizations on it, for generating \ac{gpu}-oriented
\ac{llvm}~\ac{ir}, and for performering low-level optimizations on that IR.
Through the new inferfaces, the execution of these compilation steps is
repurposed with new \ac{gpu}-specific functionality that is implemented in
\ac{gpu} extensions in the CUDAnative.jl.  For the front end, most of the
\ac{gpu}-specific functionality actually resides in the \ac{gpu} standard
library as discussed in the previous section; the front-end extensions in the
\ac{gpu} compiler are therefore minimal.

The resulting low-level, \ac{gpu}-optimized \ac{llvm}~\ac{ir} is then compiled
to \ac{ptx} by means of the \ac{llvm} NVPTX back end, which is again accessed
with the LLVM.jl wrapper package from \Cref{sec:interfaces:back-end}. This use
of an external \ac{gpu} back-end compiler rather than one embedded in the device
package diverges from the design in \Cref{img:langimpl:abstract}, as was already
hinted in \Cref{sec:vision}. For its CPU back end, the Julia compiler already
relies on CPU \ac{llvm} back ends. So any Julia distribution already includes
\ac{llvm}. The fact that \ac{llvm} can also generate excellent \ac{ptx} code for
\ac{cuda} devices when it is fed well-formed and optimized \ac{llvm} IR
code~\cite{wu2016gpucc}, voids the need for including a third-party \ac{gpu}
compiler or a reimplementation thereof in the device package. Without putting
any burden on system administrators or users to install additional packages or
tools, we can simply reuse the \ac{llvm} \ac{ptx} back end.

Before generating machine code, LLVM optimization passes extensively optimize
the \ac{llvm}~\ac{ir}. As part of that process, Julia and CUDAnative.jl lower
specific constructs and optimize certain patterns. One optimization that
drastically improves performance, is rewriting the calling convention of
entry-point functions. Semantically, Julia passes objects of an immutable type
by copying, while mutable types are passed by reference. The actual calling
convention as generated by the Julia compiler also passes aggregate immutable
types by reference, while maintaining the aforementioned semantics. In the case
of \ac{gpu} code, this means that not the aggregate argument itself, but only a
pointer to the argument will be stored in the designated state space (see
\Cref{sec:cuda:stdlib:devptr}). This space has special semantics that map well
onto typical function argument behavior ---read-only access instead of
read-write, per-kernel sharing instead of per-thread--- and typically offers
better performance than loading arguments from other memories. However, by
passing arguments by reference only the pointer will be loaded from parameter
space, and not the underlying objects. In other words, the Julia array objects
that themselves contain pointers to the actual buffers to be manipulated by the
\ac{gpu}, are not moved into designated \ac{gpu} memories to optimize
performance.

To solve this problem, we let the \ac{gpu} compiler enforce an adapted calling
convention for entry-point kernel functions: Immutable aggregates are now also
passed by value, instead of by reference. This does not change semantics as
objects of mutable types are still passed by reference. We implement this change
at the \ac{llvm}~\ac{ir} level by generating a wrapper function that takes
values as arguments, stores said values in a stack slot, and passes references
to those slots to the original entry-point function. After forced inlining and
optimization, all redundant operations disappear. Finally, the CUDAnative.jl
runtime passes all immutable arguments by value instead of by reference. This
optimization yields a speedup of up to 20\% on memory-intensive Rodinia
benchmarks, as will be discussed in \Cref{sec:evaluation:low-level}.

This optimization provides an excellent example of the code reuse enabled by our
tool flow design and the added extension interfaces. Due to that reuse, the code
to build the wrapper function and perform the necessary optimizations to inline
the code requires less than 100 lines of Julia code.

\begin{lstlisting}[float, language=C++, label={lst:vadd:cuda},
                   caption={Vector addition in \acs{cuda}~C, using the
                            \acs{cuda} run-time \acs{api}.},
                   gobble=2]
  #define cudaCall(err) // check return code for error
  #define frand() (float)rand() / (float)(RAND_MAX)

  __global__ void vadd(const float *a, const float *b,¶\newline¶                           float *c) {
    int i = blockIdx.x * blockDim.x + threadIdx.x;
    c[i] = a[i] + b[i];
  }

  const int len = 100;

  int main() {
    float *a, *b;
    a = new float[len];
    b = new float[len];
    for (int i = 0; i < len; i++) {
      a[i] = frand(); b[i] = frand();
    }

    float *d_a, *d_b, *d_c;
    cudaCall(cudaMalloc(&d_a, len * sizeof(float)));
    cudaCall(cudaMemcpy(d_a, a, len * sizeof(float),¶\newline¶           cudaMemcpyHostToDevice));
    cudaCall(cudaMalloc(&d_b, len * sizeof(float)));
    cudaCall(cudaMemcpy(d_b, b, len * sizeof(float),¶\newline¶           cudaMemcpyHostToDevice));
    cudaCall(cudaMalloc(&d_c, len * sizeof(float)));

    vadd<<<1, len>>>(d_a, d_b, d_c);

    float *c = new float[len];
    cudaCall(cudaMemcpy(c, d_c, len * sizeof(float),¶\newline¶           cudaMemcpyDeviceToHost));
    cudaCall(cudaFree(d_c));
    cudaCall(cudaFree(d_b));
    cudaCall(cudaFree(d_a));

    return 0;
  }
\end{lstlisting}

\subsection{\acs{cuda} \acs{api} Wrapper}
\label{sec:cuda:api}

The CUDAnative.jl package provides functionality related to compiling code for
\ac{cuda}~\acp{gpu}, but another important aspect of \ac{gpu} applications is to
interface directly with the device, e.g., to allocate memory, upload compiled
code, and manage execution of kernels. \ac{cuda} provides two mostly
interchangeable interfaces for this: the low-level driver \ac{api}, and the
runtime \ac{api} with higher-level semantics and automatic management of certain
resources and processes.

\begin{lstlisting}[float,language=julia, label={lst:vadd:julia},
                   caption={Vector addition in Julia using CUDAdrv.jl and
                            CUDAnative.jl.},
                   gobble=2]
  function vadd(a, b, c)
    i = (blockIdx().x-1) * blockDim().x + threadIdx().x
    c[i] = a[i] + b[i]
    return
  end

  len = 100
  a = rand(Float32, len)
  b = rand(Float32, len)

  d_a = CUDAdrv.Array(a)
  d_b = CUDAdrv.Array(b)
  d_c = similar(d_a)

  @cuda (1,len) vadd(d_a, d_b, d_c)¶\label{lst:vadd:julia:invocation}¶
  c = Base.Array(d_c)
\end{lstlisting}

\Cref{lst:vadd:cuda} shows an example vector addition in \ac{cuda}~C, using the
runtime \ac{api} to initialize and upload memory, launch the kernel, and fetch
back results. The syntax for calling kernels, \verb|vadd<<<...>>>(...)|,  hides
much of the underlying complexity: setting-up a parameter buffer, initializing
the execution configuration, acquiring a reference to the compiled kernel code,
etc.

To improve the usability of the \ac{cuda}~\ac{api} from Julia, we have created
\emph{CUDAdrv.jl}, which is available at
\url{https://github.com/JuliaGPU/CUDAdrv.jl}. This is a package wrapping the
\ac{cuda} driver \ac{api}. It offers the same level of granularity as the driver
\ac{api}, but wrapped in high-level Julia constructs for improved productivity.
Similar to the runtime \ac{api}, it automates management of resources and
processes, but always allows manual control for low-level programming tasks.
This makes the wrapper suitable for both application developers and library
programmers.

\Cref{lst:vadd:julia} shows a Julia implementation of the vector addition from
\Cref{lst:vadd:cuda}, using CUDAdrv.jl for all device interactions. It shows how
the \ac{api} wrapper vastly simplifies common operations: Memory allocation and
initialization is encoded through different constructors of the custom
\code{Array} type, \ac{api} error codes are automatically checked and converted
to descriptive exceptions, \ac{gpu} memory is automatically freed by the Julia
garbage collector, etc.

\subsection{Run-time System}
\label{sec:cuda:runtime}

While no particular attention was paid so far to the fact that the Julia
compiler is a JIT compiler, the CUDAnative.jl run-time system makes it possible
to program \acp{gpu} using dynamic programming principles, and to invoke those
programs almost at the speed of statically-compiled kernels.

Whereas calling a kernel from \acs{cuda}~C is a fully static phenomenon, our
\code{@cuda} Julia macro enables a much more dynamic approach. The \ac{gpu}
compiler is invoked, and hence kernels are compiled, upon their first
\emph{use}, i.e., right before an \code{@cuda} function call is first evaluated.
At that point, the invoked kernel and its functions are specialized and
optimized for both the active device and the run-time types of any arguments.
For additional, later occurrences of kernel invocations on arguments with
different run-time types, newly specialized and optimized code is generated.

\begin{lstlisting}[float, language=Julia, label={lst:cuda:expansion},
                   caption={Lowered code generated from the \code{@cuda}
                            invocation in \Cref{lst:vadd:julia}.},
                   gobble=2]
  # results of compile-time computations
  ## at parse time
  grid = (1,1,1)¶\label{lst:cuda:expansion:compiletime:start}¶
  block = (len,1,1)
  shmem = 0
  stream = CuDefaultStream()
  func = vadd
  args = (d_a, d_b, d_c)
  ## during type inference
  types = (CuDeviceArray{Float32,2,AS.Global},¶\newline¶         CuDeviceArray{Float32,2,AS.Global}¶\newline¶         CuDeviceArray{Float32,2,AS.Global})
  partial_key = hash(func, types)¶\label{lst:cuda:expansion:compiletime:stop}¶

  # determine the run-time environment
  age = method_age(func, $types)
  ctx = CuCurrentContext()
  key = hash(partial_key, age, ctx)

  # cached compilation
  kernel = get!(kernel_cache, key) do¶\label{lst:cuda:expansion:cache}¶
      dev = device(ctx)
      cufunction(dev, func, types)
  end

  cudacall(kernel, types, args,¶\newline¶         grid, block, shmem, stream)¶\label{lst:cuda:expansion:launch}¶
\end{lstlisting}


The specialized host code that is generated from the \code{@cuda} invocation in
\Cref{lst:vadd:julia} is shown in \Cref{lst:cuda:expansion}. Lines
\ref{lst:cuda:expansion:compiletime:start} to
\ref{lst:cuda:expansion:compiletime:stop} contain the result of compile-time
computations: Arguments to the \code{@cuda} macro are decoded during macro
expansion, and a generator function (not shown) precomputes values and
determines the kernel function signature. This signature can differ from the
types of the objects passed to \code{@cuda}, e.g., the invocation on
\cref{lst:vadd:julia:invocation} in \Cref{lst:vadd:julia} passes
\code{CUDAdrv.Array}s, but the kernel is compiled for \ac{gpu}-compatible
\code{CuDeviceArray} objects. The run-time conversion of \code{CUDAdrv.Array}
objects to their \code{CuDeviceArray} counterpart happens as part
\code{cudacall} on \cref{lst:cuda:expansion:launch}.

In addition to recompiling specialized and optimized kernels for changing
run-time types, the CUDAnative.jl runtime keeps track of the so-called
\emph{method age}, which indicates the time of definition of the function or any
of its dependents. The concept of method age is already supported in the main
Julia compiler in support of dynamic method redefinitions: Whenever a source
code fragment is edited, the containing method's age changes, and the new
version will be used for future method calls.

CUDAnative.jl also supports this concept of age. At run time, the method age and
the active \ac{cuda} context are queried. These determine whether a kernel needs
to be recompiled: A newer age indicates a redefinition of the method or any
callee, while the context determines the active device and owns the resulting
kernel object. These properties are hashed together with the type signature, and
used to query the compilation cache on \cref{lst:cuda:expansion:cache} of
\Cref{lst:cuda:expansion}. In the case of a cache miss, the kernel is compiled
and added to the cache. Finally, control is handed over to CUDAdrv.jl on
\cref{lst:cuda:expansion:launch} where \code{cudacall} converts the arguments
and launches the kernel.

The above calling sequence has been carefully optimized: Run-time operations are
avoided as much as possible, caches are used to prevent redundant computations,
code is specialized and aggressively inlined to avoid unnecessary dynamic
behavior (e.g., iterating or introspecting arguments or their types), etc. The
fast path, i.e. when no device code needs to be compiled, contains only the bare
minimum interactions with the Julia compiler and \ac{cuda} \ac{api}. As a
result, the time it takes to launch a kernel is almost equivalent to a fully
static kernel call in \acs{cuda}~C, despite all dynamic programming
capabilities. When code does need to be compiled, the time it takes to do so is
acceptably low for interactive programming purposes. Both the kernel compilation
times and their launching times will be evaluated in
\Cref{sec:evaluation:compiler}.

The support for method redefinitions with CUDAnative.jl makes it possible to
program a \ac{gpu} interactively, for example using Project~Jupyter, a popular
programming environment among scientists and teachers for programming
interactively in Julia, Python or R~\cite{ragan2014jupyter}. The environment
centers around so-called notebooks, documents that can contain both computer
code, the results from evaluating that code, and other rich-text elements. The
contents of these notebooks can be changed or evaluated in any order and at any
time, requiring a great deal of flexibility from the underlying execution
environment, e.g., to recompile code whenever it has been edited. CUDAnative.jl
makes it possible to use this highly dynamic style of programming in combination
with \acp{gpu}, for example to develop \ac{gpu} kernels by iteratively
redefining device methods and evaluating the output or performance.

This capability provides an excellent demonstration of the advantages of (i) our
vision of adding interfaces for main compiler repurposing, and (ii) our
implementation of \ac{cuda} support by means of a pure Julia device package.
This enables tight integration of \ac{gpu} support into the existing compiler,
which in turn makes the integration of \ac{gpu} support in a project like
Jupyter seamless, both for the developers of the \ac{gpu} support, and
from the perspective of Jupyter users, who get the same interactivity for host
programming and \ac{gpu} programming. All that was needed was a careful design
of the compilation cache, which was needed anyway, and 5 lines of code to
include the method age in the hashes used to access the cache.


%
%
\section{Evaluation}
\label{sec:evaluation}

To demonstrate the full capabilities of this framework, we present a three-fold
evaluation: First we evaluate the computational overhead of JIT compilation
compared to the static reference \ac{cuda}~C toolchain. Then we use a set of
standardized benchmarks to assess the run-time performance overhead of using
CUDAnative.jl for \ac{gpu} programming. The Julia implementations of these
benchmarks are ported over from \acs{cuda}~C, maintaining the low abstraction
level in order to accurately determine performance differences caused by the
programming infrastructure itself, rather than implementation choices.

The final part of the evaluation serves to illustrate the high-level programming
capabilities of the infrastructure, for which we have implemented kernels using
typical high-level programming constructs as used by Julia programmers. We
demonstrate how these constructs can be applied to \ac{gpu} programming, without
sacrificing performance.

\subsection{Experimental Set-up}
\label{sec:evaluation:setup}

\acs{cuda}~C code is compiled with the NVIDIA \ac{cuda} compiler version 8.0.61,
in combination with NVIDIA driver 375.66 and Linux 4.9.0 from Debian Stretch
(64-bit). Julia measurements are done with Julia 0.6, except for the compilation
time measurements from \Cref{sec:evaluation:compiler:codegen} where a prerelease
version of Julia 0.7 is used to reflect recent improvements in package load
times. All Julia packages in this evaluation are publicly available:
CUDAnative.jl version~0.5, CUDAdrv.jl~0.5.4 and LLVM.jl~0.5.1 using LLVM~3.9.
Our test system contains an NVIDIA GeForce GTX 1080 \ac{gpu}, two quad-core
Intel Xeon E5-2637~v2s \acsp{cpu}, and 64GB of DDR3 ECC memory.

\subsection{JIT Compiler Performance}
\label{sec:evaluation:compiler}

As described in \Cref{sec:cuda:runtime}, the CUDAnative.jl run-time system
checks before every kernel launch whether the \ac{jit} compiler needs to
generate new device code. For this to be usable in both an interactive setting,
where methods or their invocations change frequently, and in static applications
with deterministic call behavior, two metrics of compiler performance are
important: (i) the time it takes to generate \ac{ptx} code from Julia sources,
and (ii) the run-time overhead to implement the aforementioned dynamic behavior.

\subsubsection{Code Generation}
\label{sec:evaluation:compiler:codegen}

\begin{figure}[!t]
  \centering
  \includegraphics{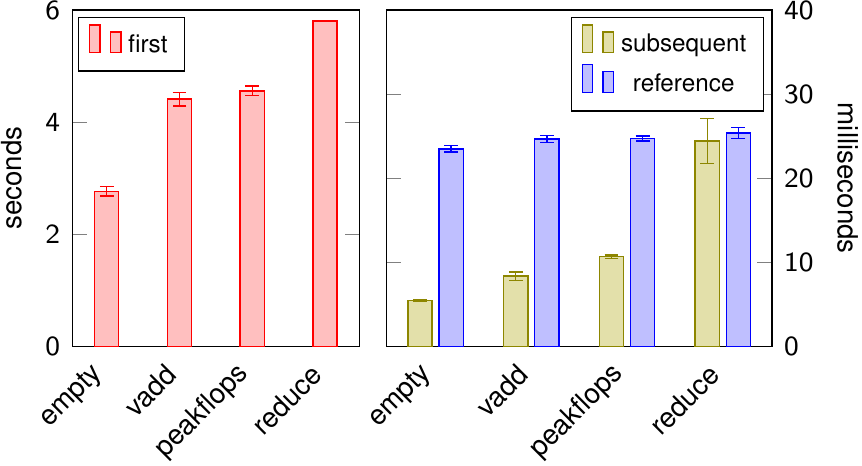}
  \vspace{-0.5em}
  \caption{Compilation times of various \ac{gpu} kernels to \ac{ptx} code. First
           and subsequent timings are taken with CUDAnative.jl, while the
           reference measurements represent compilation with the \ac{nvrtc}
           compiler library.}
  \label{img:overhead:compilation}
\end{figure}

\Cref{img:overhead:compilation} visualizes the time it takes to compile
instances of different types of kernels to \ac{ptx} code: an empty kernel, the
vector addition from \Cref{lst:vadd:julia}, a peakflops benchmark that relies on
the \ac{cuda} device library, and a high-level reduction that specializes based
on a function operator argument (that will be discussed in
\Cref{sec:evaluation:high-level}). Individual instances of these kernels are
identical but treated as unique, i.e., they do not hit the various caches in
Julia or CUDAnative.jl, but trigger a full recompilation for every invocation.
This process is purely single-threaded: neither the existing Julia compiler nor
CUDAnative.jl make use of multithreading. \Cref{img:overhead:compilation} shows
on the left how the very first \ac{gpu} compilation suffers from a significant
penalty. This can be attributed to Julia JIT-compiling CUDAnative.jl and its
dependencies for the host processor. Lowering this compilation time is an active
area of development for the Julia compiler team and is expected to improve in
future releases.

The first-time compilation penalty increases with the complexity of the kernels,
as more advanced kernels trigger the compilation of more functionality from
CUDAnative.jl. The subsequent timings on the right are much lower, and show how
(i)~Julia generates high-quality host code for CUDAnative.jl, with zero-cost
\ac{ffi} abstractions for interacting with the \ac{llvm} and \ac{cuda}
libraries, and (ii)~CUDAnative.jl has been carefully optimized to generate
device code efficiently. For example, loading and linking the \ac{cuda} device
library only happens when required by the kernel, with intermediate results
cached for future compilations.

The reference timings on the right represent the time it takes to compile
equivalent \ac{cuda}~C code. We measured this directly using the \ac{nvrtc}
library, as the \code{nvcc} compiler binary spends significant time loading
libraries and discovering a host compiler. CUDAnative.jl compares favorably
against the \ac{cuda}~C toolchain, but exhibits worse scaling behavior as Julia
source code is considerably more complex to analyze~\cite{nash2016inference}. As
mentioned before, compilation times are expected to improve with future Julia
versions. Furthermore, we believe that the relatively low complexity of typical
GPU applications will mask this behavior.

\subsubsection{Kernel Launch Overhead}
\label{sec:evaluation:compiler:launch}

\begin{table}[!t]
  \sffamily
  \renewcommand{\arraystretch}{1.2}
  \caption[]{GPU and CPU execution times for an empty kernel, measuring
             respectively time between \acs{cuda} events surrounding the kernel,
             and wall-clock time to record these events and launch the kernel.}
  \label{tbl:overhead:execution}
  \centering
  \vspace{-0.5em}
  \begin{tabular}{|c|c c|}
    \hline
                  & GPU time                     & CPU time                       \\
    \hline
    \acs{cuda} C  & \SI{6.12(71)}{\micro\second} & \SI{12.49(148)}{\micro\second} \\
    CUDAdrv.jl    & \SI{6.97(70)}{\micro\second} & \SI{12.50(80)}{\micro\second}  \\
    CUDAnative.jl & \SI{8.17(82)}{\micro\second} & \SI{13.77(97)}{\micro\second}  \\
    \hline
  \end{tabular}
\end{table}

With statically compiled \ac{cuda}~C code, the run-time cost of launching a
kernel is dominated entirely by the \ac{cuda} libraries and underlying hardware.
In the case of CUDAnative.jl, \Cref{sec:cuda:runtime} described how launching a
kernel entails many more tasks with the goal of a highly dynamic programming
environment: converting arguments, (re)compiling code, instantiating a \ac{cuda}
module and function object, etc. These code paths have been carefully optimized
to avoid run-time overhead as much as possible.

To determine this overhead, we launch an empty kernel and measure the elapsed
execution time, both on the GPU using \ac{cuda} events and on the CPU using
regular wall-clock timers. \Cref{tbl:overhead:execution} shows these
measurements for statically-compiled C code using the \ac{cuda} driver \ac{api},
Julia code performing the exact same static operations with CUDAdrv.jl, and
dynamic Julia code using CUDAnative.jl to compile and execute the empty kernel.
Neither of the GPU and CPU time measurements show significant overhead when only
using CUDAdrv.jl, thanks to Julia's excellent generated code quality and
zero-cost \ac{ffi}. With CUDAnative.jl, which internally uses CUDAdrv.jl,
minimal overhead is introduced by the check for the current method age. We
consider this negligible: In the case of realistic kernels it is dwarfed by the
time to copy the kernel parameter buffer to the device.

\subsection{Low-level Kernel Programming}
\label{sec:evaluation:low-level}

To asses the raw performance of \ac{gpu} kernels written in Julia, we have
ported several \acs{cuda}~C benchmarks from the Rodinia benchmark suite to
CUDAnative.jl~\cite{che2009rodinia}. These ports are available at
\url{https://github.com/JuliaParallel/rodinia/}. Porting \ac{gpu} code is a
non-trivial effort, so we have focused on the smallest benchmarks of the suite,
taking into account use of \ac{gpu} features that are or were not yet supported
by the CUDAnative.jl infrastructure, such as constant memory.

\subsubsection{Lines of Code}

\begin{figure}[t!]
  \centering
  \includegraphics{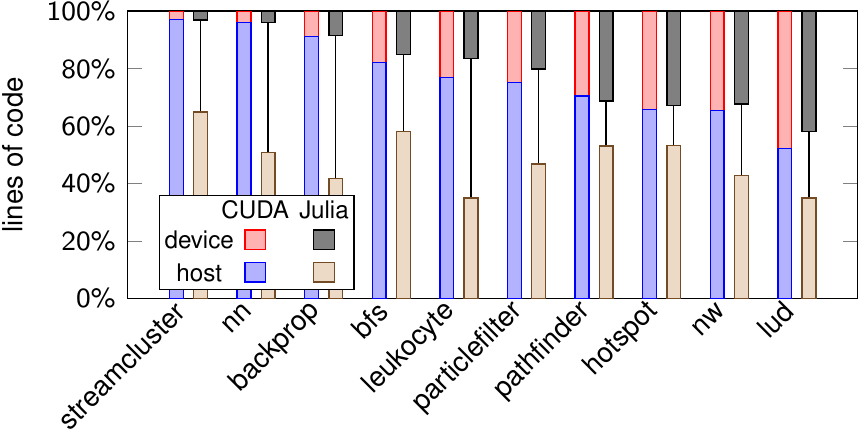}
  \vspace{-0.5em}
  \caption{Lines of host and device code of selected Rodinia benchmarks,
           normalized against the total LOC of their \acs{cuda}~C
           implementations. On average, the Julia - CUDAnative.jl versions are
           32\% shorter: device code is reduced by 8\%, while the amount of host
           code is reduced by 38\%.}
  \label{img:sloc}
\end{figure}

To accurately determine performance differences that result from using the Julia
language and the CUDAnative.jl compiler, we maintained the low-level semantics
of the original \acs{cuda}~C kernel code, whereas for the host code we applied
high-level programming concepts and used the \ac{api} wrapper from
\Cref{sec:cuda:api}. This is clearly visible in \Cref{img:sloc}, which shows
significant reductions in lines of code for the host part of the benchmarks,
while the amount of device code decreases much less. Even so, this low-level
style of Julia still significantly improves the programming experience, with,
e.g., dynamic types, checked arithmetic, an improved programming environment,
etc.

\subsubsection{Kernel Performance}

\begin{figure}[!t]
  \centering
  \includegraphics{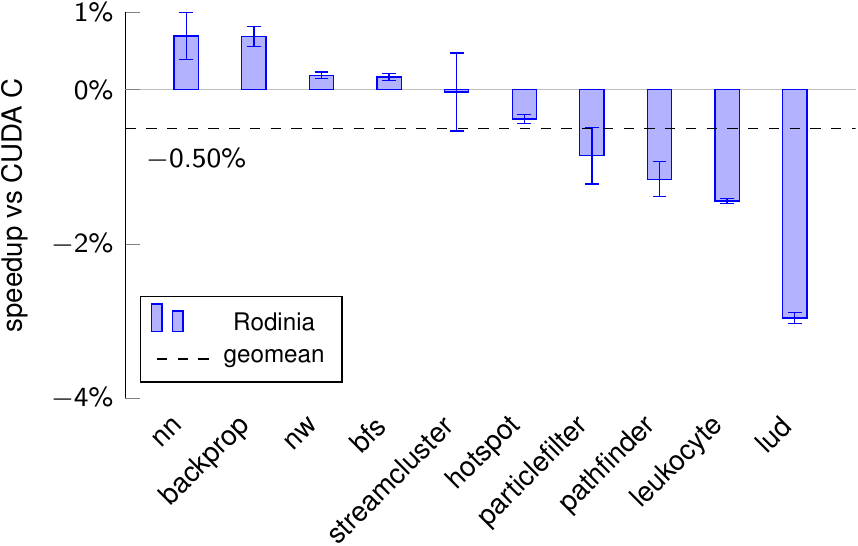}
  \vspace{-1em}
  \caption{Kernel performance of CUDAnative.jl vs \acs{cuda}~C.}
  \label{img:perf}
\end{figure}

\Cref{img:perf} visualizes the performance difference between kernels
implemented in Julia using CUDAnative.jl and their reference Rodinia
\acs{cuda}~C implementations. The results are obtained by running the
\acs{cuda}~C and Julia versions of each benchmark multiple times, measuring the
execution times of every \ac{gpu} kernel using the \code{nvprof} profiling tool
from the \ac{cuda} toolkit. Memory transfers and other host-side operations are
excluded from these timings. The mean execution time per kernel is then
estimated by maximum-likelihood fitting a lognormal
distribution~\cite{mashey2004benchmarks, ciemiewicz2001mean}, and the total
execution time of each benchmark is the sum of its kernels' mean execution
times. \Cref{img:perf} shows the speedup of the thus measured Julia total
execution time over the \ac{cuda}~C total execution time, with error margins
determined by propagating the distributions' variances across
operations~\cite{giordano2016uncertainty}. The average speedup is computed as
the geometric mean, visualized by a dashed horizontal line.

On average, we measure a slowdown of $0.50\%$ compared to \acs{cuda}~C kernels
compiled with \code{nvcc}. This is close to the relative speedup of $0.8\%$ as
achieved by \code{gpucc} on a wider range of Rodinia
benchmarks~\cite{wu2016gpucc}. As \code{gpucc} is built on the same \ac{llvm}
back end as CUDAnative.jl, we can conclude that using Julia for low-level
\ac{gpu} kernel programming does not incur a substantial slowdown.

\subsection{High-level \ac{gpu} Programming}
\label{sec:evaluation:high-level}

\begin{span}
  \begin{lstlisting}[language=julia, label={lst:evaluation:broadcast},
                     caption={Using CuArrays.jl to perform a fused broadcast
                              on the \acs{gpu}.},
                     gobble=4]
    X = CuArray(rand(42))¶\label{line:broadcast:alloc}¶
    f(x) = 3x^2 + 5x + 2¶\label{line:broadcast:def}¶
    Y = f.(2 .* X.^2 .+ 6 .* X.^3 .- sqrt.(X))¶\label{line:broadcast:call}¶
  \end{lstlisting}

  \begin{lstlisting}[language=CUDAC, label={lst:evaluation:broadcast_equiv},
                     caption={Equivalent \acs{cuda}~C kernel code for the fused
                              broadcast in \Cref{lst:evaluation:broadcast}.
                              Host code has been omitted for brevity, and
                              would be similar to \Cref{lst:vadd:cuda}.},
                     gobble=4]
    __device__ double f(double x) {
      return 3*pow(x, 2) + 5*x + 2;
    }
    __global__ void kernel(double *X, double *Y, int N) {
      int i = blockIdx.x * blockDim.x + threadIdx.x;
      if (i < N) Y[i] = f(2*X[i]+6*pow(X[i],3)-sqrt(X[i]));
    }
  \end{lstlisting}
\end{span}

To demonstrate the high-level programming potential of this infrastructure, we
use \emph{CuArrays.jl}~\cite{innes2017cuarrays} (not to be confused with the
lightweight array type provided by CUDAdrv.jl). This package defines an array
type for data that lives on the \ac{gpu}, but exposes host-level operations that
are implemented using the infrastructure from this paper to execute on the
device. For example, constructing a \code{CuArray} object will allocate data on
the \ac{gpu} using CUDAdrv.jl, adding two such host objects together will queue
an addition kernel on the \ac{gpu} using CUDAnative.jl, etc. Certain other
functionality is implemented using optimized host libraries like cuBLAS or
cuDNN, but that is not relevant to the work in this paper.

The example from \Cref{lst:evaluation:broadcast} shows how to load the
CuArrays.jl package, generate input data and upload it to the \ac{gpu} on
\cref{line:broadcast:alloc}, defining an auxiliary function for the sake of this
example on \cref{line:broadcast:def}, and finally a series of element-wise
operations including a call to the newly defined function on
\cref{line:broadcast:call}. These operations, prefixed by a dot to indicate the
element-wise application, are syntactically fused together into a single
broadcast operation~\cite{johnson2017dots}:
\lstinline{Y = broadcast(x -> f(2x^2+6x^3-sqrt(x)), X)} where the first argument
is a lambda containing the fused operations from \cref{line:broadcast:call}. The
implementation of \code{broadcast} in CuArrays.jl then compiles this function
using CUDAnative.jl, inlining calls to both the lambda and underlying function
\code{f}.

Conceptually, broadcasting a function over \ac{gpu} arrays like \code{CuArray}
is straightforward: each thread processes a single element, the grid can be
constructed relatively naively, there are no cross-thread dependencies, etc.
However, the actual implementation relies on several advanced properties of the
Julia language and compiler. For one, Julia specializes functions on the types
of its arguments (which includes the shape of each container). This makes it
possible to write generic code, nonetheless compiled to statically typed
assembly without type checks. Furthermore, every function in Julia has its own
type. This permits use of higher-order arguments, even user-defined ones as in
\Cref{lst:evaluation:broadcast}, that still result in specialized code without,
e.g., indirect function calls or calls to the runtime. In fact, the \ac{ptx}
code generated from \Cref{lst:evaluation:broadcast} is identical to that
generated from the equivalent \ac{cuda}~C code of
\Cref{lst:evaluation:broadcast_equiv}, with the exception of slightly different
inlining decisions made by the various compilers. The amount of source code,
however, is dramatically reduced: Kernels can be expressed much more naturally,
and \ac{api} interactions (not shown in \Cref{lst:evaluation:broadcast_equiv}
for the sake of brevity) disappear for most use cases.

\begin{lstlisting}[float, language=Julia, label={lst:evaluation:reduce},
                   caption={Reducing an array of custom objects on the
                            \acs{gpu}.},
                   gobble=2]
  # illustrational type that implements addition
  struct Point{T}
      x::T
      y::T
  end
  +(a::Point{T}, b::Point{T}) where {T} =
    Point(a.x+b.x, a.y+b.y)

  data = [Point(rand(Int64)%100, rand(Int64)%100)¶\newline¶        for _ in 1:42]
  X = CuArray(data)
  Y = reduce(+, #=neutral element=# Point(0,0), X)¶\label{lst:evaluation:invocation}¶
\end{lstlisting}

Where \code{broadcast} is a relatively simple operation, the CuArrays.jl package
also implements other data processing algorithms optimized for \acp{gpu}. One
noteworthy such algorithm is \code{reduce}, with a parallel implementation based
on shuffle instructions~\cite{nvidia2015shuffle}. These provide a
means to exchange data between threads within the same thread block, without
using shared memory or having to synchronize execution.  The implementation of
shuffle in CUDAnative.jl exposes a fully generic interface that specializes on
the argument types, whereas even the official \ac{cuda}~C intrinsics are limited
to certain primitive types. As a result, reducing a \code{CuArray} offers the
same flexibility as described for \code{broadcast}.
\Cref{lst:evaluation:reduce} demonstrates reducing an array of custom objects
using the \code{+} operator (more complex operators are supported but would make
the example more confusing). Again, the invocation on
\cref{lst:evaluation:invocation} compiles to a single kernel specialized on each
of the arguments to \code{reduce}. This specialization includes generating
sequences of 32-bit shuffle instructions to move the 128-bit \code|Point{Int64}|
objects between threads, courtesy of a generated function producing
\ac{llvm}~\ac{ir} with LLVM.jl as explained in \Cref{sec:interfaces}.  The final
abstraction completely hides this complexity, however, and demonstrates how
metaprogramming can be used to selectively override function behavior and use
LLVM.jl to tap into the full potential of the underlying compiler.


The abstractions from this section are idiomatic Julia code, made possible by
the CUDAnative.jl \ac{jit} compiler. CuArrays.jl demonstrates the power of such
abstractions, by combining the convenience of a host library containing
predefined abstractions, with the flexibility and performance of manual device
programming. This greatly improves programmer productivity, while offering the
necessary expressiveness to target, e.g., \acp{gpu} or other parallel
accelerators.

\section{Related Work}
\label{sec:related}

In recent times, many developments have added support for \acp{gpu} and similar
accelerators to general purpose, high-level languages without depending on a
lower-level, device specific language such as \ac{cuda} or OpenCL. One popular
approach is to host a \ac{dsl} in the general-purpose language, with properties
that allow it to compile more easily for, e.g., \acp{gpu}. For example,
Accelerate defines an embedded array language in
Haskell~\cite{chakravarty2011accelerate}, while Copperhead works with a
functional, data-parallel subset of Python~\cite{catanzaro2011copperhead}.
Parakeet uses a similar Python subset, with less emphasis on the functional
aspect~\cite{rubinsteyn2012parakeet}, whereas PyGPU specializes its \ac{dsl} for
image processing algorithms~\cite{lejdfors2007pygpu}. Other research defines
entirely new languages, such as Lime~\cite{dubach2012lime},
Chestnut~\cite{stromme2012chestnut} or
HIPA\textsuperscript{cc}~\cite{membarth2016hipa}. In each of these cases, the
user needs to gain explicit knowledge about this language, lowering his
productivity and impeding reuse of existing code.

Our work proposes compiler infrastructure that allows writing code for
accelerators directly in the high-level source language, tightly integrated with
the main compiler and language ecosystem. Rootbeer targets similar
programmability with Java, but requires manual build-time actions to
post-process and compile kernel source code~\cite{pratt2012rootbeer}. Jacc
features automatic run-time compilation and extraction of implicit parallelism,
but requires the programmer to construct manually an execution task-graph using
a relatively heavy-weight \ac{api}~\cite{clarkson2015jacc}.

By extending the main compiler, we greatly reduce the effort required to support
new targets. This type of extensible programming has been extensively researched
in the past~\cite{solntseff1974survey}, and has seen a recent
revival~\cite{zingaro2007modern}, but to our knowledge has not focused on
extensibility of compiler processes for the purpose of targeting new hardware
and environments with minimal code duplication. The Rust language has
experimental support for NVIDIA~\acp{gpu} that does reuse low-level \ac{llvm}
infrastructure, but lacks integration with the higher levels of the compiler and
focuses on statically compiling device code with little run-time interactions or
optimizations~\cite{holk2013rustgpu}. NumbaPro, being a Python compiler, does
target a much higher-level language and interface, with corresponding run-time
interactions like \ac{jit} compilation based on the kernel type
signature~\cite{lam2015numba}. However, it uses a custom Python compiler which
significantly complicates the implementation and is not fully compatible with
the Python language specification.

\section{Conclusion and Future Work}
\label{sec:future}

\paragraph*{\textbf{Conclusion}}

We presented an approach for efficiently adding support for new hardware or
other environments to an existing programming language. We proposed a set of
interfaces to repurpose the existing compiler, while maximizing reuse of
functionality. We implemented these interfaces in the compiler for the
high-level Julia programming language, and used that infrastructure to add
support for NVIDIA \acp{gpu}. We then used the Rodinia benchmark suite to show
how Julia can be used to write \ac{gpu} code that performs similar to
\ac{cuda}~C. By integrating with the existing compiler, code compatibility is
improved and many existing Julia packages can be used on the \ac{gpu} without
extra effort.

Our work on CUDAnative.jl makes it possible to apply high-level principles to
\ac{gpu} programming, for example dynamically typed kernels or interactive
programming tools like Jupyter. Furthermore, CUDAnative.jl and its
\ac{gpu} \ac{jit} compiler make it possible to create highly flexible,
high-level abstractions. The CuArrays.jl package demonstrates this, with a
interface that combines the convenience of host-level libraries with the
flexibility of manual device programming.

\paragraph*{\textbf{Status}}

Our changes to the Julia compiler's interfaces have been accepted, and are part
of the 0.6 stable release. The extension interfaces are already being used by
other researchers and developers to add support for more platforms to the Julia
compiler, from similar hardware like AMD GPUs, to WebAssembly for targeting web
browsers. These new developments invariably make use of LLVM.jl, and often mimic
the design of CUDAnative.jl as a starting point.

Our packages are critical to large parts of the Julia GPU infrastructure for
NVIDIA GPUs. Furthermore, they work out
of the box with Jupyter, enabling interactive programming and effective use of
\ac{gpu} hardware.

\paragraph*{\textbf{Future Work}}

We plan to improve support for \ac{gpu} hardware features, and create high-level
abstractions that maintain the ability to express low-level behavior. This
includes a unified approach to \ac{gpu} memory types, idiomatic support for
communication primitives, etc. We are also working on compiler improvements to
enable even more powerful abstractions, for example contextual method dispatch
based on run-time device properties. This can both enhance expressiveness of the
abstractions, and improve performance of the generated code.

\section*{Acknowledgments}
\label{sec:acks}

This work is supported by the Institute for the Promotion of Innovation by
Science and Technology in Flanders (IWT Vlaanderen), and by Ghent University
through the Concerted Research Action on distributed smart cameras.

%

\bibliographystyle{IEEEtran}
\bibliography{main}

%

\begin{IEEEbiography}[{\includegraphics[height=1.25in]{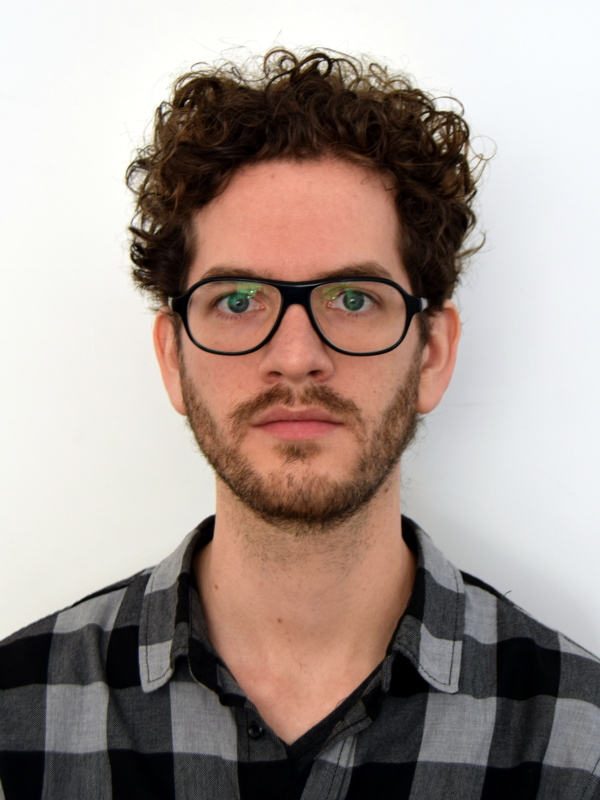}}]
{Tim Besard} is a PhD student at Ghent University in the Computer Systems Lab.
He obtained his MSc in Computer Engineering from University College Ghent in
2011. His research focuses on compilation techniques of high-level languages for
GPUs.
\end{IEEEbiography}

\begin{IEEEbiography}[{\includegraphics[height=1.25in]{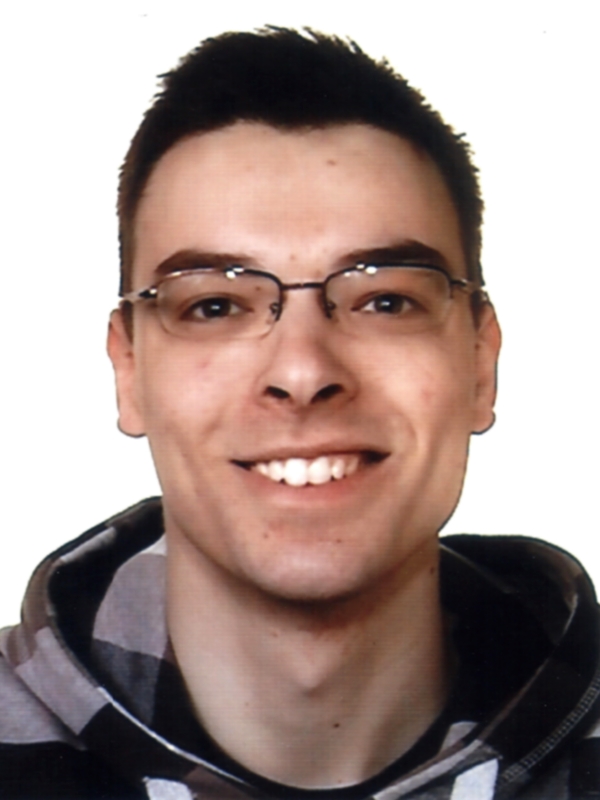}}]
{Christophe Foket} is a post-doctoral researcher at Ghent University in the
Computer Systems Lab. He obtained his MSc and PhD degrees in Computer Science
from Ghent University's Faculty of Engineering in 2009 and 2015. His research
focuses on compilation techniques of high-level languages for GPUs.
\end{IEEEbiography}

\begin{IEEEbiography}[{\includegraphics[height=1.25in]{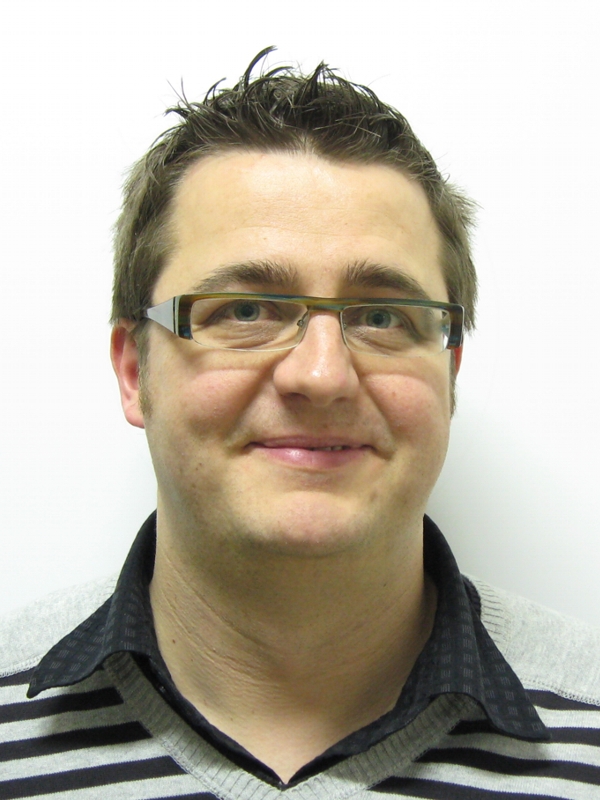}}]
{Bjorn De Sutter} is associate professor at Ghent University in the Computer
Systems Lab. He obtained his MSc and PhD degrees in Computer Science from Ghent
University's Faculty of Engineering in 1997 and 2002. His research focuses on
the use of compiler techniques to aid programmers with non-functional aspects of
their software, such as performance, code size, reliability, and software
protection.
\end{IEEEbiography}

\end{document}